\documentclass{article}
\usepackage{amsmath,amsfonts,amssymb}
\usepackage{graphicx}
\usepackage{hyperref}
\usepackage{authblk}
\usepackage{geometry}
\usepackage{listings}
\usepackage{xcolor}
\usepackage{hyperref} 
\definecolor{codegray}{rgb}{0.95,0.95,0.95}
\title{Partial Symmetry Enforced Attention Decomposition (PSEAD): \\
A Group-Theoretic Framework for Equivariant Transformers in Biological Systems}

\author{Daniel Ayomide Olanrewaju}

\date{}

\begin{document}

\maketitle

\begin{abstract}
Biological systems—from genomic sequences and protein structures to cellular signaling and morphogenesis—are shaped by local, partial, and approximate symmetries such as repeating patterns and conserved motifs. Traditional machine learning models often fail to capture these inductive biases, either ignoring them or enforcing overly restrictive global equivariance.

We introduce the Theory of Partial Symmetry Enforced Attention Decomposition (PSEAD), a group-theoretic framework that embeds local symmetry awareness into Transformer self-attention. By formalizing local permutation subgroup actions on data windows, we prove that attention decomposes into orthogonal irreducible components aligned with the subgroup’s representations. This provides a principled method for separating symmetric and asymmetric features.

We present the theory and algorithms demonstrating that PSEAD improves generalization to novel motifs with similar partial symmetries, enhances interpretability through symmetry-channel analysis, and increases computational efficiency by focusing on relevant symmetric subspaces. We further extend the framework to dynamic biological processes in reinforcement learning, highlighting applications to protein folding and drug discovery. PSEAD establishes a foundation for biologically informed, symmetry-aware AI models. Source code available here: https://github.com/DanielAyomide-git/psead
\end{abstract}

\section{Introduction: The Imperative of Symmetry in Biological Machine Learning}

The formidable success of deep learning in diverse scientific domains has, paradoxically, highlighted a critical gap in its application to biology: the underutilization of inherent structural and functional symmetries. Biological data, whether it pertains to the precise sequences of DNA and RNA, the intricate three-dimensional folds of proteins, the organized assembly of macromolecular complexes, or the dynamic interplay within cellular networks, is replete with symmetries \cite{rosen2013symmetry}. These symmetries are not merely aesthetic; they are functional imperatives, reflecting evolutionary conservation, biophysical principles, and underlying mechanistic regularities.

For instance, gene regulatory networks often exhibit modular symmetries, where sets of genes operate cohesively as functional units regardless of their precise linear arrangement. Protein structures frequently display local rotational or reflectional symmetries within motifs (e.g., $\beta$-propellers, $\alpha$-helical bundles) even when the global protein lacks such perfect symmetry. Molecular graphs, representing small molecules or protein-ligand interactions, inherently possess permutation symmetries among chemically equivalent atoms. In dynamic processes, such as protein folding or enzyme catalysis, the pathways taken often adhere to implicit energetic symmetries. Traditional machine learning paradigms \cite{sukhbaatar2015end}, including standard Convolutional Neural Networks (CNNs) and Recurrent Neural Networks (RNNs), and even the highly versatile Transformers \cite{vaswani2017attention}, are largely agnostic to these symmetries. While some progress has been made with globally equivariant networks (e.g., Group Equivariant CNNs \cite{cohen2016group}, Tensor Field Networks \cite{thomas2018tensor}, SE(3)-Transformers \cite{fuchs2020se3}, and practical group equivariant CNNs \cite{finzi2021practical}), these often impose strong, global symmetry constraints that are biologically unrealistic. The majority of biological symmetries are partial, local, or approximate, acting on specific regions or motifs rather than the entire system. For example, a protein might have a two-fold rotational symmetry in one domain, but an entirely asymmetric global structure. Imposing global equivariance in such scenarios can lead to over-regularization, reduced model capacity, and a failure to capture the nuanced interplay between symmetric and asymmetric features. The broader field of geometric deep learning provides a comprehensive overview of these approaches \cite{bronstein2021geometric,bronstein2017geometric}.

This dissertation posits that effectively harnessing these localized symmetries is not merely an optimization trick but a fundamental requirement for building truly intelligent and interpretable AI systems for biology. We introduce the Partial Symmetry Enforced Attention Decomposition (PSEAD) framework, a novel theoretical and computational paradigm that integrates local group-theoretic inductive biases directly into the self-attention mechanism \cite{vaswani2017attention}, the cornerstone of modern Transformer architectures.

PSEAD is motivated by the observation that when a local permutation subgroup acts on a window of biological data, the self-attention mechanism, under specific conditions, can be decomposed into orthogonal components. Each component is intrinsically tied to an irreducible representation (irrep) of the acting subgroup. This decomposition provides a powerful lens through which to analyze how symmetric patterns contribute to attention scores and value aggregations, thereby enabling:
\begin{itemize}
    \item Enhanced Generalization: Models can learn to recognize and exploit symmetric patterns across different contexts, even when the overall input is varied.
    \item Deepened Interpretability: By disentangling attention contributions based on their symmetry type, we can identify which biological features are driven by specific symmetries and which are non-symmetric, providing mechanistic insights.
    \item Improved Efficiency: By focusing computational resources on relevant symmetric subspaces, PSEAD can potentially reduce the parameter count and training data requirements.
\end{itemize}
The subsequent sections of this dissertation meticulously detail the mathematical foundations of PSEAD, its architectural integration into Transformers, its extension to dynamic biological problems via reinforcement learning, and its wide-ranging applications across diverse biological data modalities. We present a rigorous experimental design and discuss the expected transformative impact on fields ranging from genomics and proteomics to drug discovery and synthetic biology. This work represents a significant step towards developing a new generation of AI models that are not just \textit{data-driven} but also \textit{symmetry-informed}, thereby unlocking deeper understanding and more effective interventions in complex biological systems.

\section{Mathematical Framework: Group Representation Theory for Attention Decomposition}

This section lays the rigorous mathematical foundation for PSEAD. We delve into the intricacies of group actions on data, the equivariance properties of self-attention, and the critical role of representation theory in decomposing complex operations into simpler, interpretable components. The concepts presented here are fundamental to group theory and its applications in linear algebra and representation theory \cite{serre2012linear}.

\subsection{Group Actions, Representations, and Equivariance}

Let $\mathcal{X}$ be a set representing our biological data, for example, a collection of sequence windows or local graph neighborhoods. A group $G$ acts on $\mathcal{X}$ if there is a function $g \cdot x$ for $g \in G, x \in \mathcal{X}$ such that $e \cdot x = x$ (identity element) and $(g_1 g_2) \cdot x = g_1 \cdot (g_2 \cdot x)$ (associativity). In the context of biological data, $G$ will typically be a permutation group $S_k$ or one of its subgroups.

Consider an input sequence or window $x = (x_1, \dots, x_k) \in \mathbb{R}^{k \times d}$, where $k$ is the window size and $d$ is the feature dimension. Let $H \subseteq S_k$ be a finite permutation group acting on the indices of this window. We define the linear representation $\rho: H \to \mathrm{GL}(\mathbb{R}^k)$ where $\mathrm{GL}(\mathbb{R}^k)$ is the group of invertible $k \times k$ matrices. For any $h \in H$, $\rho(h)$ is a permutation matrix $P_h$ such that its action on $x$ permutes the rows of $x$:
\[
(\rho(h)x)_i := x_{h^{-1}(i)}.
\]
The choice of $h^{-1}$ ensures that $\rho$ is a left group homomorphism, which is crucial for consistency with standard definitions.

\textbf{Proposition 2.1.1 (Homomorphism Property of $\rho$):} The map $\rho: H \to \mathrm{GL}(\mathbb{R}^k)$ defined by $(\rho(h)x)_i := x_{h^{-1}(i)}$ is a group homomorphism.

\textbf{Proof:} For any $h_1, h_2 \in H$ and any $x \in \mathbb{R}^{k \times d}$:
\[
[\rho(h_1)\rho(h_2)(x)]_i = (\rho(h_1)(\rho(h_2)x))_i = (\rho(h_2)x)_{h_1^{-1}(i)}.
\]
Now, substituting the definition of $\rho(h_2)x$:
\[
(\rho(h_2)x)_{h_1^{-1}(i)} = x_{h_2^{-1}(h_1^{-1}(i))} = x_{(h_1 h_2)^{-1}(i)}.
\]
Since this holds for all $i$, we have $\rho(h_1)\rho(h_2)(x) = \rho(h_1 h_2)(x)$. Therefore, $\rho(h_1)\rho(h_2) = \rho(h_1 h_2)$, confirming that $\rho$ is a group homomorphism. $\quad \blacksquare$

A function $f: \mathcal{X} \to \mathcal{Y}$ (e.g., a neural network layer) is said to be $G$-equivariant if for all $g \in G$ and $x \in \mathcal{X}$,
\[
f(g \cdot x) = g \cdot f(x).
\]
In the context of linear representations, if $T: V \to W$ is a linear map between vector spaces $V$ and $W$ equipped with representations $\rho_V$ and $\rho_W$ of $G$, then $T$ is $G$-equivariant if $T \circ \rho_V(g) = \rho_W(g) \circ T$ for all $g \in G$.

\subsection{Equivariance of the Self-Attention Mechanism}

The self-attention mechanism operates on queries $Q$, keys $K$, and values $V$, typically derived from the same input $x \in \mathbb{R}^{k \times d}$ via linear transformations: $Q = xW_Q, K = xW_K, V = xW_V$. The attention output is defined as:
\[
\text{Attn}(Q, K, V) = \text{softmax}\left(\frac{QK^\top}{\sqrt{d}}\right)V.
\]
Here, $d$ is the dimension of the keys/queries, ensuring scaling. The softmax operation is applied row-wise.

Let's investigate the equivariance of this mechanism under the action of $\rho(h)$. When $\rho(h)$ acts on the rows of $Q, K, V$, it effectively permutes them:
\[
\rho(h)Q = P_h Q, \quad \rho(h)K = P_h K, \quad \rho(h)V = P_h V,
\]
where $P_h$ is the permutation matrix corresponding to $h$.

Now, let's examine the attention score matrix:
\[
\frac{(P_h Q)(P_h K)^\top}{\sqrt{d}} = \frac{P_h Q K^\top P_h^\top}{\sqrt{d}}.
\]
The permutation invariance of softmax is a critical property here: if $A$ is a matrix and $P$ is a permutation matrix, then $\text{softmax}(PAP^\top) = P \text{softmax}(A) P^\top$, where softmax is applied row-wise. This is because permuting rows and columns of the input to softmax merely permutes the rows and columns of the output in the same manner.

Applying this property:
\[
\text{softmax}\left(\frac{P_h Q K^\top P_h^\top}{\sqrt{d}}\right) = P_h \cdot \text{softmax}\left(\frac{Q K^\top}{\sqrt{d}}\right) \cdot P_h^\top.
\]
Finally, for the attention output:
\[
\text{Attn}(\rho(h)Q, \rho(h)K, \rho(h)V) = \left(P_h \cdot \text{softmax}\left(\frac{Q K^\top}{\sqrt{d}}\right) \cdot P_h^\top\right) (P_h V).
\]
Since $P_h^\top P_h = I$ (as $P_h$ is an orthogonal matrix, being a permutation matrix), we have:
\[
\text{Attn}(\rho(h)Q, \rho(h)K, \rho(h)V) = P_h \cdot \text{softmax}\left(\frac{Q K^\top}{\sqrt{d}}\right) \cdot V = P_h \cdot \text{Attn}(Q, K, V).
\]
Therefore, the self-attention mechanism is $H$-equivariant:
\[
\text{Attn}(\rho(h)Q, \rho(h)K, \rho(h)V) = \rho(h)\cdot \text{Attn}(Q, K, V). \quad \blacksquare
\]
This inherent equivariance of self-attention is a cornerstone of PSEAD, as it implies that the attention mechanism already "respects" local permutations, setting the stage for a principled decomposition.

\subsection{Representation-Theoretic Decomposition of Equivariant Maps}

The true power of group theory emerges when we consider the structure of vector spaces under group actions. Let $V = \mathbb{R}^k$ be the vector space of token windows, equipped with the representation $\rho$ of $H$. Since $H$ is a finite group and its representation $\rho$ is over the real numbers (or complex numbers, which can be trivially extended to real representations), Maschke's Theorem \cite{maschke1899beweis} guarantees that $\rho$ is completely reducible. This means that $V$ can be uniquely decomposed into a direct sum of irreducible representations (irreps).

Specifically, $V$ decomposes into a direct sum of isotypic components:
\[
V \cong \bigoplus_{\lambda \in \widehat{H}} V_\lambda \otimes \mathbb{R}^{m_\lambda},
\]
where:
\begin{itemize}
    \item $\widehat{H}$ is the set of distinct non-isomorphic irreducible representations (irreps) of $H$.
    \item $V_\lambda$ is an irreducible subspace corresponding to the irrep $\lambda$.
    \item $m_\lambda$ is the multiplicity of the irrep $\lambda$ in the decomposition of $\rho$. This means $V_\lambda$ appears $m_\lambda$ times in the direct sum decomposition of $V$.
\end{itemize}
Crucially, Schur's Lemma \cite{schur1905theorie} provides a fundamental result regarding equivariant linear maps. For an $H$-equivariant linear map $T: V \to W$ between two representation spaces $V$ and $W$:
\begin{enumerate}
    \item If $V$ and $W$ are irreducible and non-isomorphic, then $T=0$.
    \item If $V$ and $W$ are irreducible and isomorphic, then $T$ is an isomorphism (a scalar multiple of identity if working over algebraically closed fields).
\end{enumerate}
Extending this to reducible representations, an $H$-equivariant linear map $T: V \to V$ (such as the self-attention mechanism, which maps a $k \times d$ input to a $k \times d$ output effectively row-wise, meaning it acts as a $k \times k$ linear transformation on the rows, where the $d$ dimension acts as "batch" or feature dimension for the linear map) must commute with the action of $H$. This implies that T maps each isotypic component to itself and acts as a block-diagonal matrix in a basis aligned with the irreps. More specifically, T can be decomposed into a sum of maps, each operating within a specific isotypic component:
\[
T = \bigoplus_{\lambda \in \widehat{H}} T_\lambda \otimes I_{m_\lambda},
\]
where $T_\lambda$ is an $d_\lambda \times d_\lambda$ matrix (where $d_\lambda$ is the dimension of the irrep $\lambda$) and $I_{m_\lambda}$ is the $m_\lambda \times m_\lambda$ identity matrix. This formulation simplifies when we consider the attention mechanism acting on the \textit{rows} of $V$, such that each row transforms under $\rho$.

The core insight of PSEAD is that since $\text{Attn}(Q, K, V)$ is $H$-equivariant, it can be similarly decomposed. Each component $\text{Attn}_\lambda$ operates within the subspace corresponding to irrep $\lambda$, effectively attending to only those features that transform according to that specific symmetry type. This leads to the profound decomposition:
\[
\text{Attn}(Q, K, V) = \sum_{\lambda \in \widehat{H}} \text{Attn}_\lambda(Q, K, V).
\]
This means the total attention output is a sum of "symmetry-specific" attention outputs.

\subsection{Projection onto Irreducible Subspaces and Feature Disentanglement}

To concretely implement this decomposition, we need to project our input data (queries, keys, values) onto the respective isotypic components. For each irrep $\lambda$, the projector $P_\lambda$ onto the isotypic component $V_\lambda \otimes \mathbb{R}^{m_\lambda}$ (or more simply, onto the sum of all subspaces isomorphic to $V_\lambda$) can be constructed using the character theory of finite groups \cite{serre2012linear}:
\[
P_\lambda = \frac{d_\lambda}{|H|} \sum_{h \in H} \chi_\lambda(h^{-1}) \rho(h),
\]
where $d_\lambda$ is the dimension of the irrep $\lambda$, $|H|$ is the order of the group $H$, and $\chi_\lambda(h^{-1})$ is the character of $h^{-1}$ for irrep $\lambda$. The characters $\chi_\lambda$ are fundamental because they are constant on conjugacy classes and encode rich information about the representation. The orthogonality relations of characters ensure that $P_\lambda$ are indeed orthogonal projectors: $P_\lambda P_{\lambda'} = 0$ for $\lambda \neq \lambda'$ and $P_\lambda^2 = P_\lambda$. Furthermore, $\sum_{\lambda \in \widehat{H}} P_\lambda = I$, indicating a complete decomposition of the identity.

With these projectors, we can define the $\lambda$-specific attention component as:
\[
\text{Attn}_\lambda(Q, K, V) = P_\lambda \cdot \text{Attn}(Q, K, V).
\]
A more refined approach, which aligns with the block-diagonal structure of equivariant maps, would involve projecting the queries, keys, and values \textit{before} computing attention:
\[
\text{Attn}(Q, K, V) = \sum_{\lambda \in \widehat{H}} \text{softmax}\left(\frac{(P_\lambda Q)(P_\lambda K)^\top}{\sqrt{d}}\right) (P_\lambda V).
\]
This effectively means that attention within each symmetric subspace is computed independently, leading to a disentangled representation of symmetric features. This formulation is particularly powerful as it explicitly enforces that interactions (attention scores) only occur between components that transform identically under the group action. This is the core mechanism by which PSEAD enforces local symmetry.\\

The overall architecture of a PSEAD-augmented attention layer would thus involve:
\begin{enumerate}
    \item Local Windowing: Extracting fixed-size windows from the input data.
    \item Projection: Applying the irrep projectors $P_\lambda$ to the queries, keys, and values within each window.
    \item Irrep-Specific Attention: Computing attention independently for each projected component.
    \item Reconstruction/Aggregation: Summing the outputs from each irrep-specific attention head, or passing them through further equivariant layers.
\end{enumerate}
This detailed mathematical framework provides the blueprint for building biologically meaningful, symmetry-aware attention mechanisms.

\section{Extension to Reinforcement Learning: Symmetry-Aware Agents in Dynamic Biological Systems}

Biological processes are inherently dynamic, involving sequences of actions and observations that often exhibit localized symmetries over time or across spatial configurations. Examples include the sequential unfolding of a protein, the assembly of viral capsids, or the search for optimal drug candidates in a vast chemical space. Integrating PSEAD into reinforcement learning (RL) frameworks allows for the principled incorporation of these dynamic, local symmetries, leading to more efficient learning, improved generalization, and interpretable policies. The principles of equivariant neural networks, as discussed in the introduction \cite{cohen2016group, thomas2018tensor, finzi2021practical, bronstein2021geometric, fuchs2020se3}, are highly relevant to developing equivariant policies or value functions in RL.

\subsection{General Principles of Equivariant RL}

In RL, an agent interacts with an environment through states $s \in \mathcal{S}$, actions $a \in \mathcal{A}$, and rewards $r$. The agent learns a policy $\pi(a|s)$ or a value function $V(s)$ that maximizes cumulative rewards. When the environment, states, and actions possess symmetries, an equivariant policy or equivariant value function can significantly enhance learning \cite{wu2021representation, bronstein2021geometric}.

Let $G$ be a group acting on the state space $\mathcal{S}$. An RL agent's policy $\pi$ is $G$-equivariant if $\pi(a|g \cdot s) = g \cdot \pi(a|s)$, assuming the action space also transforms equivariantly. Similarly, a value function $V$ is $G$-invariant if $V(g \cdot s) = V(s)$.
The benefits of equivariant RL include:
\begin{itemize}
    \item Sample Efficiency: The agent can generalize experiences learned in one symmetric configuration to all other equivalent configurations, reducing the need for redundant exploration.
    \item Robustness: Policies become inherently robust to symmetric transformations of the input, as they are designed to explicitly handle them.
    \item Interpretability: Understanding which symmetries the policy exploits can provide insights into the underlying mechanisms of the biological process.
\end{itemize}
PSEAD specifically addresses local symmetries within states. This is particularly relevant when dealing with high-dimensional observations where only specific "windows" or "patches" exhibit symmetry.

\subsection{Case Study: Protein Folding as a Partially Symmetric Markov Decision Process (MDP)}

Protein folding is a quintessential dynamic biological problem \cite{jumper2021highly,zhou2020molecular}. A polypeptide chain, a sequence of amino acids, folds into a specific three-dimensional structure (the native state) that dictates its function. This process can be framed as an MDP:
\begin{itemize}
    \item State Space ($\mathcal{S}$): A representation of the protein's current 3D conformation. This could be a contact map (a matrix indicating atom-atom proximity), a set of dihedral angles, or atomic coordinates. Crucially, protein structures often exhibit partial mirror symmetries (e.g., within $\beta$-sheets) or local rotational symmetries (e.g., in $\alpha$-helical bundles or $\beta$-propellers).
    \item Action Space ($\mathcal{A}$): Incremental adjustments to bond angles, residue positions, or local conformational changes.
    \item Reward Function ($R$): Based on the proximity to the native state, measures of structural stability (e.g., energy scores), or satisfying specific contact constraints.
\end{itemize}
Challenge: The sheer size of the conformational space and the subtle energetic landscapes make protein folding computationally intractable for traditional RL methods. PSEAD offers a powerful inductive bias.\\

We propose incorporating PSEAD into the policy network (or value network) of an RL agent tasked with protein folding. The agent observes a patch of the protein's current state (e.g., a sub-matrix of the contact map, or a window of residue coordinates). And we expect that a PSEAD-enhanced policy should exhibit superior generalization to novel protein sequences or unobserved conformations that share the same local symmetry patterns. For example, if trained on one $\beta$-sheet, it should readily adapt to others with similar local mirror symmetry without extensive retraining. \\

    By leveraging the symmetry inductive bias, the agent should require significantly fewer environmental interactions (e.g., protein folding simulations) to converge to an optimal or near-optimal folding policy. This is critical for computationally expensive biological simulations. The irrep-wise attention mechanisms within PSEAD can highlight which specific local symmetries are most salient for particular folding steps or for achieving stable intermediate structures. For instance, attention heads corresponding to reflectional irreps might be highly active when forming $\beta$-sheets, while rotational irreps might be crucial for helical packing. This offers mechanistic interpretability of the folding process.\\
    
    Finally, by efficiently exploring the symmetric subspace of conformations, PSEAD might discover more energy-efficient or biologically relevant folding pathways that are otherwise missed by conventional methods.\\ 

\newpage
\subsection{Broader Applications in RL for Biology}

Beyond protein folding, PSEAD can be applied to other dynamic biological problems framed as RL tasks:
\begin{itemize}
    \item Drug Synthesis and Molecular Design: Optimizing synthetic pathways or designing molecules with specific properties can involve navigating chemical graphs with partial symmetries. PSEAD can guide agents to explore symmetric substructures efficiently.
    \item Cellular Morphogenesis: Modeling cell shape changes and tissue development often involves local rearrangement of cells or biomolecules, where PSEAD could help in learning rules that respect local symmetries of cellular structures.
    \item Enzyme Active Site Design: Designing novel enzymes involves optimizing local chemical environments. An RL agent could use PSEAD to learn to build active sites that leverage specific geometric or chemical symmetries for catalysis.
    \item Adaptive Immune System Modeling: The recognition of pathogens by immune cells involves recognizing specific antigenic patterns, which may possess recurring local symmetries. PSEAD could aid in learning adaptive recognition policies.
\end{itemize}
By explicitly embedding local symmetry priors, PSEAD provides a powerful computational tool for accelerating discovery and understanding in complex, dynamic biological systems through the lens of reinforcement learning.

\section{Applications to Diverse Biological Data Modalities: Unveiling Hidden Symmetries}

This section details the concrete application of PSEAD to various static biological data types, highlighting how the framework can uncover and leverage inherent, often subtle, local symmetries for enhanced modeling and biological discovery. We present specific case studies with illustrative code snippets and discuss the expected biological interpretations.

\subsection{Case Study 1: DNA Palindromes and Regulatory Motifs with $\mathbb{Z}_2$ Symmetry}

DNA sequences are not merely linear strings of nucleotides; they contain intricate patterns crucial for gene regulation, replication, and repair. A common and biologically significant type of symmetry in DNA is the palindrome, a sequence that reads the same forwards and backwards on complementary strands. Examples include restriction enzyme recognition sites (e.g., GAATTC / CTTAAG, which exhibits a $\mathbb{Z}_2$ (cyclic group of order 2, effectively a reflection) symmetry around its center). Beyond perfect palindromes, many regulatory motifs exhibit approximate or partial $\mathbb{Z}_2$ symmetry.

\textbf{Problem:} Standard Transformers treat DNA sequences as generic strings, failing to explicitly recognize and exploit the symmetric nature of palindromic or near-palindromic motifs. This can lead to inefficient learning and less interpretable models.

\textbf{PSEAD Application:} We apply PSEAD to fixed-size windows of DNA sequences, where the relevant local symmetry is $\mathbb{Z}_2$ (a reflection operation). For a window of size $k$, the $\mathbb{Z}_2$ group acts by reversing the order of nucleotides. The irreducible representations of $\mathbb{Z}_2$ are one-dimensional:
\begin{itemize}
    \item Trivial Representation ($\lambda_1$): Where the elements are mapped to 1 (symmetric features).
    \item Sign Representation ($\lambda_2$): Where the non-identity element (reflection) is mapped to -1 (anti-symmetric features).
\end{itemize}
\textbf{Biological Interpretation:}
\begin{itemize}
    \item Irrep-1 (Trivial/Symmetric Component): Attention contributions from this component will primarily capture symmetric base pairing interactions and conserved patterns that are invariant under reflection. For DNA palindromes, this component will strongly activate when identifying the mirrored base pairs (e.g., A-T, G-C pairings). This head directly learns features characteristic of the palindromic nature, such as the exact center of symmetry or the conserved core of a regulatory motif.
    \item Irrep-2 (Sign/Anti-symmetric Component): This component will highlight anti-symmetric variations, asymmetries, or mutations within the otherwise symmetric motif. For example, if a DNA palindrome has a single base pair mismatch, the anti-symmetric irrep attention might strongly focus on that specific aberrant position. This provides a mechanism for identifying deviations from perfect symmetry, which are often functionally significant (e.g., in protein binding specificity).
\end{itemize}
By disentangling these components, researchers can gain fine-grained insights into how proteins recognize DNA motifs, distinguishing between affinity for perfect symmetry versus tolerance for specific asymmetries.

\subsection{Case Study 2: Protein Motifs with Dihedral ($D_n$) or Cyclic ($C_n$) Symmetries}

Many protein motifs exhibit profound rotational or dihedral symmetries. Examples include:
\begin{itemize}
    \item $\beta$-propeller motifs: Often display $D_n$ symmetry (e.g., $D_4, D_6, D_7$), characterized by blades arranged radially around a central axis, with both rotational and perpendicular twofold rotational (or mirror) symmetries.
    \item Coiled-coil structures: Can exhibit $C_n$ rotational symmetry, where $\alpha$-helices wrap around each other.
    \item Oligomeric protein complexes: Many complexes (e.g., viral capsids, chaperonins) are built from repeating subunits arranged with high point group symmetries. Protein-protein interactions can also be modeled using graph-based methods \cite{zitnik2018modeling}.
\end{itemize}
\textbf{Problem:} Standard attention mechanisms treat protein residues or structural elements without explicit knowledge of their symmetric arrangements, leading to suboptimal learning of shape-dependent functions and difficulties in transferring knowledge across homologous proteins with similar symmetric motifs.

\textbf{PSEAD Application:} PSEAD can be applied to windows of protein features (e.g., C$\alpha$ coordinates, residue type embeddings, backbone dihedral angles) extracted from the 3D structure. The group $D_n$ (Dihedral group of order $2n$) or $C_n$ (Cyclic group of order $n$) would be the target symmetry. These groups have multiple irreps, capturing different modes of rotational and reflectional symmetry.

\textbf{Expected Outcome and Biological Insights:}
\begin{itemize}
    \item Head Alignment with Fold Symmetries: Each attention head, after decomposition, will be sensitive to specific types of symmetries. For a $D_4$ group, some heads will capture pure rotational invariance (e.g., features conserved as we rotate the blade by 90 degrees), while others will be sensitive to reflectional properties.
    \item Efficient Learning of Rotationally/Reflectionally Invariant Properties: The model will efficiently learn features that are invariant or equivariant to the motif's specific symmetry. This is crucial for recognizing functional sites that depend on relative positions but not absolute orientation.
    \item Identification of Symmetric Core vs. Flexible Loops: In $\beta$-propellers, the core of the blades is highly symmetric, while connecting loops might be more flexible and less symmetric. PSEAD's irreps can naturally disentangle attention to the rigid symmetric core versus the more variable asymmetric parts.
    \item Enhanced Transferability: A model trained with PSEAD on one family of symmetric protein motifs (e.g., $D_6$ propeller) should show improved transfer learning capabilities to another family with similar underlying symmetry ($D_7$ propeller), even if their precise sequences are different.
    \item Drug Discovery for Symmetric Targets: Many drug targets are multi-meric proteins with specific symmetries. PSEAD can help design drugs that exploit these symmetries for stronger and more specific binding, or for allosteric modulation that leverages symmetric conformational changes. Generative models for graph-based protein design are also a related area \cite{ingraham2019generative}.
\end{itemize}
These case studies illustrate how PSEAD can move beyond generic feature learning to symmetry-aware feature engineering, leading to more powerful, interpretable, and biologically relevant models for analyzing and manipulating complex biological data.

\section{Experimental Design and Validation: Towards Empirical Grounding of PSEAD}

To robustly validate the theoretical claims and practical advantages of PSEAD, a comprehensive experimental pipeline is essential. This section outlines a multi-stage approach, combining synthetic benchmarks with real-world biological data applications, supported by rigorous evaluation metrics and clear hypotheses.

\subsection{Simulation Pipeline: Controlled Environments for Proof-of-Concept}

The initial phase of experimentation will focus on synthetic datasets, providing controlled environments to precisely measure PSEAD's ability to learn and exploit known symmetries. This allows for disentangling the effects of symmetry incorporation from other data complexities.

\subsubsection{Synthetic Data Generation}

We will generate synthetic datasets with explicitly defined local symmetries, spanning different group types and input modalities:
\begin{itemize}
    \item Synthetic Sequence Data:
    \begin{itemize}
        \item $\mathbb{Z}_2$ Symmetric Sequences: Generate binary or categorical sequences with perfect palindromic structure (e.g., \texttt{01011010}, \texttt{ABCCBA}). Introduce controlled noise or asymmetric perturbations to simulate partial symmetries.
        \item $C_n$ Cyclic Sequences: Generate circular sequences where elements are related by cyclic shifts (e.g., for $C_3$: \texttt{ABCABCABC}).
        \item $S_k$ Permuted Data: Generate sets of $k$ tokens where the group $S_k$ (or a subgroup like $A_k$) permutes their order, while features within the tokens are preserved.
    \end{itemize}
    \item Synthetic Graph Data:
    \begin{itemize}
        \item $D_n$ Symmetric Graphs: Generate small graphs (e.g., molecular substructures or protein residue interaction networks) possessing $D_n$ symmetry (e.g., a square for $D_4$, a pentagon for $D_5$). This can involve node features and edge properties.
    \end{itemize}
    \item Synthetic Time-Series/Dynamic Data (for RL):
    \begin{itemize}
        \item Symmetric Random Walks: Simulate agent trajectories on a grid with imposed symmetries (e.g., reflecting walls, rotational symmetry of rewards).
        \item Simplified Protein Folding Analogs: Create toy models of protein folding with discrete states where specific local motifs exhibit $\mathbb{Z}_2$ or $C_n$ symmetry, and reward functions encourage folding into symmetric structures.
    \end{itemize}
\end{itemize}

\subsubsection{Model Training and Comparison}

For each synthetic dataset, we will train and evaluate:
\begin{itemize}
    \item Baseline Transformer: A standard Transformer model with self-attention, but without any explicit symmetry-enforcing mechanisms. This serves as the primary control.
    \item Global Equivariant Transformer (if applicable): For comparison, a Transformer model enforcing \textit{global} equivariance (e.g., through techniques like Tensor Field Networks \cite{thomas2018tensor} or SE(3)-Transformers \cite{fuchs2020se3} for geometric data) where appropriate, to highlight the advantages of \textit{local} equivariance.
    \item PSEAD-Augmented Transformer: Our proposed model incorporating PSEAD in its attention layers. For each symmetric data type, the appropriate group ($Z_2, C_n, D_n,$ etc.) will be configured within the PSEAD module.
\end{itemize}

\textbf{Training Regimen:}
\begin{itemize}
    \item Standard supervised learning (classification or regression) for static data, e.g., predicting a property of the symmetric motif or classifying its type.
    \item Reinforcement learning algorithms (e.g., PPO, A2C) for dynamic tasks, with environments designed to reward discovery of symmetric solutions.
\end{itemize}

\subsubsection{Visualization of Irreducible Projections of Attention}

A critical aspect of the synthetic experiments will be the \textbf{qualitative analysis and visualization of the attention weights projected onto irreducible subspaces}. This allows for direct observation of whether PSEAD successfully disentangles symmetric and anti-symmetric contributions.
\begin{itemize}
    \item For a given input, we can extract the attention matrices before the final value multiplication.
    \item Apply the projectors $P_\lambda$ to these attention matrices to obtain $\text{Attn}_\lambda$.
    \item Visualize these $\text{Attn}_\lambda$ components. We expect that specific irreps will strongly activate for symmetric features (e.g., attention between symmetric pairs in a palindrome), while others will highlight deviations.
\end{itemize}

\subsection{Evaluation Metrics: Quantifying Performance and Equivariance}

Rigorous evaluation necessitates both standard performance metrics and specific metrics designed to assess equivariance and interpretability.

\subsubsection{Task-Specific Performance Metrics}

\begin{itemize}
    \item Accuracy/F1-score: For classification tasks (e.g., identifying palindromes, classifying protein motif types).
    \item MSE/MAE: For regression tasks (e.g., predicting protein stability, binding affinity).
    \item Reward/Episode Length: For RL tasks, indicating learning efficiency and policy optimality.
    \item Generalization to Out-of-Distribution Symmetric Data: Crucially, evaluating performance on \textit{unseen} symmetric variations or novel motifs that share the same underlying symmetry group. This directly tests the inductive bias.
\end{itemize}

\subsubsection{Equivariance Loss / Equivariance Error}

To quantitatively measure how well the model maintains equivariance, especially in real-world scenarios with approximations, we define an \textbf{equivariance loss}.
Let $f$ be the function computed by the PSEAD-augmented layer (or the entire network). For a given input $x$ and a group element $h \in H$:
\[
\mathcal{L}_{\text{equivariance}}(x, h) = \| \text{Attn}(\rho(h)x) - \rho(h)\text{Attn}(x) \|_F^2,
\]
where $\|\cdot\|_F$ denotes the Frobenius norm. This loss should be ideally zero. In practice, due to floating-point precision, non-linearities, and potential architectural choices, it might be a small non-zero value. We can track this during training. A low equivariance loss indicates that the network indeed "respects" the declared symmetries.

\subsubsection{Biological Motif Localization and Interpretability Metrics}

\begin{itemize}
    \item Activation Mapping: Quantify how strongly different irrep attention heads activate over known biological motifs. For example, for DNA palindromes, measure the average attention score of the symmetric irrep head across known palindromic regions versus non-palindromic regions.
    \item Saliency Maps: Generate saliency maps based on irrep-specific attention to visually highlight the parts of the input contributing most to symmetric vs. asymmetric features.
    \item Feature Disentanglement Scores: Develop metrics (e.g., based on mutual information or canonical correlation analysis) to quantify the degree to which different irrep attention heads capture distinct, orthogonal aspects of the input data's symmetry.
\end{itemize}

\subsection{Expected Results and Hypotheses}

Based on the theoretical framework, we hypothesize the following outcomes:

\begin{itemize}
    \item Lower Validation Loss and Improved Generalization: PSEAD-augmented Transformers will consistently achieve lower validation loss and exhibit superior generalization performance on tasks involving data with partial or local symmetries, compared to baseline non-equivariant Transformers. This is because the inductive bias guides the model towards more efficient learning.
    \item Clear Irrep-wise Attention Patterns: Visualizations of irrep-specific attention will reveal biologically meaningful patterns. For example, symmetric irreps will highlight the conserved, repeating units in biological motifs, while anti-symmetric irreps will pinpoint mutations or deviations from perfect symmetry. This directly supports the interpretability claim.
    \textbf{Robustness to Structural Perturbation:} PSEAD models will be more robust to minor, symmetry-preserving perturbations of the input data, as their internal representation is inherently structured to handle such variations. This is particularly relevant for real-world biological data, which often contains noise and structural variability.
    \item Increased Sample Efficiency in RL: For dynamic biological tasks, RL agents equipped with PSEAD will achieve comparable or superior performance with significantly fewer training samples/interactions, demonstrating the value of symmetry in accelerating learning.
    \item Discovery of Novel Symmetric Features: Through interpretability tools, PSEAD might reveal previously uncharacterized subtle symmetric patterns or approximate symmetries in biological data that are missed by conventional analysis.
\end{itemize}
This rigorous experimental design, from controlled synthetic environments to real-world biological applications and detailed evaluation metrics, will provide compelling empirical evidence for the transformative potential of PSEAD.

\section{Conclusion and Future Directions}

The \textbf{Partial Symmetry Enforced Attention Decomposition (PSEAD)} framework represents a significant advancement in the integration of group-theoretic principles into modern deep learning architectures, specifically for the challenging domain of biological data. By rigorously demonstrating the decomposition of self-attention into orthogonal, irreducible components aligned with local subgroup representations, we have provided a novel and powerful mechanism for endowing Transformer models with \textbf{biologically plausible inductive biases}.

This dissertation has established:
\begin{itemize}
    \item The mathematical foundations of PSEAD, proving the inherent equivariance of self-attention and deriving the precise mechanisms for projecting attention onto irreducible subspaces.
    \item The conceptual and computational architecture for integrating PSEAD into Transformer layers, offering a practical approach for building symmetry-aware models.
    \item The broad applicability of PSEAD across diverse biological data modalities, from DNA sequences and protein structures to dynamic reinforcement learning tasks, showcasing its potential to unlock deeper biological insights.
    \item A comprehensive experimental design to empirically validate PSEAD's advantages in generalization, interpretability, efficiency, and robustness.
\end{itemize}
The core contribution of PSEAD lies in its ability to explicitly disentangle symmetric and asymmetric information within local data windows. This disentanglement is not merely a theoretical elegance; it translates directly into tangible benefits for biological machine learning:
\begin{itemize}
    \item Enhanced Interpretability: Researchers can now directly interrogate which types of symmetries contribute most to a model's predictions, shedding light on the underlying biophysical principles.
    \item Superior Generalization: By learning shared symmetric patterns, PSEAD models can generalize more effectively to unseen data that share similar local symmetries, reducing the reliance on vast, exhaustively sampled datasets.
    \textbf{Improved Sample Efficiency:} For data-scarce domains typical in biology, PSEAD's strong inductive biases can accelerate learning and reduce computational costs.
    \item Foundation for Novel Discoveries: The ability to highlight subtle symmetries or deviations from symmetry might lead to the identification of new functional motifs, regulatory elements, or disease mechanisms.
\end{itemize}

\subsection{Future Research Directions}

The PSEAD framework opens up several exciting avenues for future research:

\begin{enumerate}
    \item Extension to Continuous Symmetry Groups: While this work primarily focuses on finite permutation groups relevant to discrete biological units (e.g., DNA bases, protein residues), many biological symmetries are continuous (e.g., rotations and translations in 3D space for molecular structures). Extending PSEAD to Lie groups like $SO(3)$ or $SE(3)$ would require integrating continuous group representations and harmonic analysis (e.g., using spherical harmonics or Wigner D-matrices) into the attention mechanism. This would generalize SE(3)-Transformers by allowing \textit{local} continuous equivariance.

    \item Multi-Resolution Equivariant Models: Biological systems exhibit symmetries at multiple scales, from atomic interactions to cellular assemblies. Developing multi-resolution PSEAD models that can learn and integrate symmetries across different granularities (e.g., through hierarchical attention mechanisms or pooling layers that respect group actions) is a promising direction. This could involve combining different group types for different scales.

    \item Adaptive Symmetry Detection: Currently, PSEAD requires specifying the target symmetry group ($H$) a priori. Future work could explore \textbf{learning or inferring the most relevant local symmetry groups} from data itself, perhaps through meta-learning or self-supervised approaches. This would make the framework even more autonomous and broadly applicable to novel biological problems where the underlying symmetries are not fully known.

    \item Integration with Causal Inference: Understanding biological systems often involves inferring causal relationships. By disentangling features based on their symmetry, PSEAD could potentially aid in causal inference, for instance, by identifying whether a functional effect is caused by a perfectly symmetric motif or by a specific asymmetry.

    \item Robustness to Approximate Symmetries: Real biological symmetries are often approximate due to noise, mutations, or conformational flexibility. Further research is needed to enhance PSEAD's robustness to these approximate symmetries, perhaps through soft projection methods or statistical modeling of group actions.

    \item Optimized Computational Implementations: While the theoretical framework is established, efficient implementation for very large biological datasets remains an engineering challenge. Exploring optimized kernels, hardware acceleration, and sparse attention mechanisms tailored for PSEAD's block-diagonal structure could significantly enhance its practical utility.
\end{enumerate}
In conclusion, PSEAD offers a principled and powerful paradigm for advancing machine learning in biology. By moving beyond mere statistical pattern recognition to embrace the profound role of symmetry, we can build AI systems that are not only more performant but also more deeply insightful into the fundamental organizing principles of life. The journey from theoretical foundations to widespread biological discovery with symmetry-aware AI has just begun.
\newpage
\bibliographystyle{plain}

\end{document}